\documentclass[a4, amssymb,amsmath,aps,prl,twocolumn,superscriptaddress,showpacs]{revtex4-1}
\usepackage{amssymb,amsmath}
\usepackage{color}
\usepackage{graphicx}
\usepackage{epstopdf}

\newcommand{\srsl}{Sr$_2$IrO$_4$}
\newcommand{\srbl}{Sr$_3$Ir$_2$O$_7$}
\newcommand{\lacuo}{La$_2$CuO$_4$}

\begin{document}
\title{The importance of $XY$ anisotropy in Sr$_2$IrO$_4$ revealed by magnetic critical scattering experiments}

\author{J. G. Vale}
\email{j.vale@ucl.ac.uk}
\affiliation{London Centre for Nanotechnology and Department of Physics and Astronomy, University College London, Gower Street, London, WC1E 6BT, United Kingdom}
\affiliation{Laboratory for Quantum Magnetism, Ecole Polytechnique F\'ed\'erale de Lausanne (EPFL), CH-1015, Switzerland}

\author{S. Boseggia}
\affiliation{London Centre for Nanotechnology and Department of Physics and Astronomy, University College London, Gower Street, London, WC1E 6BT, United Kingdom}
\affiliation{Diamond Light Source Ltd, Diamond House, Harwell Science and
Innovation Campus, Didcot, Oxfordshire, OX11 0DE, United Kingdom}

\author{H. C. Walker}
\affiliation{Deutsches Elektronen Synchrotron DESY, 22607 Hamburg, Germany}
\affiliation{ISIS Neutron and Muon Source, Rutherford Appleton Laboratory, Didcot, Oxfordshire, OX11 0QX, United Kingdom}

\author{R. Springell}
\affiliation{Royal Commission for the Exhibition of 1851 Research Fellow, Interface Analysis Centre, School of Physics, University of Bristol, Bristol, BS2 8BS, United Kingdom}

\author{Z. Feng}
\affiliation{London Centre for Nanotechnology and Department of Physics and Astronomy, University College London, Gower Street, London, WC1E 6BT, United Kingdom}

\author{E. C. Hunter}
\affiliation{School of Physics and Astronomy, The University of Edinburgh, James Clerk Maxwell Building, Mayfield Road, Edinburgh, EH9 2TT, United Kingdom}

\author{R. S. Perry}
\affiliation{London Centre for Nanotechnology and Department of Physics and Astronomy, University College London, Gower Street, London, WC1E 6BT, United Kingdom}

\author{D. Prabhakaran}
\affiliation{Clarendon Laboratory, Department of Physics, University of Oxford, Parks Road, Oxford, OX1 3PU, United Kingdom}

\author{A. T. Boothroyd}
\affiliation{Clarendon Laboratory, Department of Physics, University of Oxford, Parks Road, Oxford, OX1 3PU, United Kingdom}

\author{S. P. Collins}
\affiliation{Diamond Light Source Ltd, Diamond House, Harwell Science and
Innovation Campus, Didcot, Oxfordshire, OX11 0DE, United Kingdom}

\author{H. M. R\o{}nnow}
\affiliation{Laboratory for Quantum Magnetism, Ecole Polytechnique F\'ed\'erale de Lausanne (EPFL), CH-1015, Switzerland}
\affiliation{Neutron Science Laboratory, Institute for Solid State Physics (ISSP), University of Tokyo, Kashiwa, Chiba 277-8581, Japan}

\author{D. F. McMorrow}
\affiliation{London Centre for Nanotechnology and Department of Physics and Astronomy, University College London, Gower Street, London, WC1E 6BT, United Kingdom}

\date{\today}

\begin{abstract}
The magnetic critical scattering in  \srsl\ has been characterized using X-ray resonant magnetic scattering (XRMS) both below and above the 3D antiferromagnetic ordering temperature, $T_N$.
The order parameter critical exponent below $T_{\text{N}}$ is found to be $\beta$=0.195(4), in the range of the 2D $XYh_4$ universality class. Over an extended temperature range above $T_{\text{N}}$, the amplitude and correlation length of the intrinsic critical fluctuations are well described by the 2D Heisenberg model with $XY$ anisotropy. This contrasts with an earlier study of the 
critical scattering  over a more limited range of temperature which found agreement with the theory of the isotropic 2D Heisenberg quantum antiferromagnet, developed to describe the critical fluctuations of the conventional Mott insulator La$_2$CuO$_4$ and related systems.
Our study therefore establishes the importance of $XY$ anisotropy in the low-energy effective Hamiltonian of \srsl, the prototypical spin-orbit Mott insulator. 
\end{abstract}

\maketitle

The Ruddlesden-Popper series Sr$_{n+1}$Ir$_{n}$O$_{3n+1}$ of perovskite iridates has emerged as a fruitful arena in which to explore the effects of electron correlations in the strong spin-orbit coupling limit. The first two members of this series, single-layer \srsl\ ($n=1$) and bi-layer \srbl\ ($n=2$), are believed to exemplify a new class of spin-orbit Mott insulators. Of central importance to our understanding of these materials is the emergence of a $j_{\text{eff}}$=1/2 groundstate by the combined action of a strong cubic crystal-field and spin orbit interactions on the $5d^5$ electrons of the Ir$^{4+}$ ions \cite{kim2009}. The weakened electron correlations typical of the $5d$ elements then splits the $j_{\text{eff}}$=1/2 band, opening a gap, leading to a Mott-like state.  

\srsl\ in particular has attracted considerable attention because of its striking similarities to La$_2$CuO$_4$ in terms of both its structural and magnetic properties. The magnetic structures and excitations of \srsl\ have been investigated in a number of X-ray resonant magnetic scattering (XRMS) studies \cite{kim2009, kim2012_sr214, boseggia2013_prl, moretti2014, kim2014, liu2015, olalde_OK} which have allowed an effective low-energy Hamiltonian to be proposed and refined.
\srsl\ forms an antiferromagnetic structure below $T_{\text{N}}$ $\!\sim\!225$~K in which the moments are confined to the $\mathbf{a}$-$\mathbf{b}$ planes and canted to follow rigidly the correlated rotation of the oxygen octahedra of the $I4_1/acd$ crystal structure \cite{boseggia2013_jpcm}. A resonant inelastic X-ray scattering (RIXS) experiment \cite{kim2012_sr214} has revealed a dispersion relation somewhat reminiscent of that displayed by \lacuo\ albeit with a lower energy scale and much stronger further neighbour couplings, which can be derived from a smaller ratio of on-site repulsion over hopping amplitude \cite{dallapiazza2012}. This result suggests that the low-energy isospin dynamics of the $j_{\text{eff}}$=1/2 states in \srsl\ may, to leading order, be mapped onto an effective isotropic two-dimensional Heisenberg Hamiltonian in agreement with predictions  by Jackeli and Khaliullin \cite{jackeli2009}. 

Critical scattering studies provide information complementary to that obtainable from the ordered state. For thermally driven transitions in classical systems, issues such as dimensionality, relevant anisotropies, etc., can be addressed by determining the critical exponents both below and above the transition temperature \cite{collins1989}. One question that naturally arises is:  in what ways, if any, do the critical fluctuations of a lattice decorated by $j_{\text{eff}}$=1/2 isospins differ from the case of S=1/2 spins? 
For \lacuo\ and related 2D Cu$^{2+}$ (S=1/2) systems it has been established that the instantaneous magnetic scattering function S($\mathbf{Q}$) is well described by 
the critical properties of the 2D quantum S=1/2 Heisenberg antiferromagnet on a square-lattice (2DQHAFSL)  \cite{manousakis1991, keimer1992, ronnow1999, ronnow2001, caretta2000}. The 2DQHAFSL model itself has been extensively studied using a range of theoretical and computational techniques, with results that are in broad agreement within the range of applicability of the assumptions used \cite{chakravarty1988, makivic1991, hasenfratz1991, cuccoli1998}. 
Here for the sake of brevity we compare our data to the results obtained by quantum Monte Carlo (QMC) techniques \cite{makivic1991}. 

Fujiyama {\it et al.}~\cite{fujiyama2012} have measured the critical magnetic scattering from \srsl\ using XRMS in the interval
$T_{\text{N}}$ to $T_{\text{N}}\!+\!25$ K. Their principal results are that the 
critical fluctuations are 2D over the entire temperature interval investigated, with an evolution of the in-plane correlation length consistent with that expected for the 2DQHAFSL model \cite{chakravarty1988, makivic1991}.
The latter was interpreted by Fujiyama {\it et al.}~as evidence in favour of the theoretical proposition by Jackeli and Khaliullin \cite{jackeli2009} that the full isospin Hamiltonian (including anisotropies other than that due to Hund's coupling) can be mapped onto an effective, isotropic Heisenberg Hamiltonian. However, the good agreement reported between data and predictions of the 2DQHAFSL model required a value of the nearest-neighbour exchange coupling of $J\!=\!100(10)$ meV which they deduced by fitting their data. A subsequent RIXS study \cite{kim2012_sr214} of the one-magnon dispersion in \srsl\ provided the value $J\!=\!60$ meV, and a next-nearest-neighbour ferromagnetic coupling $J^\prime\!=\!-20$ meV, seemingly difficult to reconcile with the data of Fujiyama {\it et al.} 

\begin{figure}[t]
\includegraphics{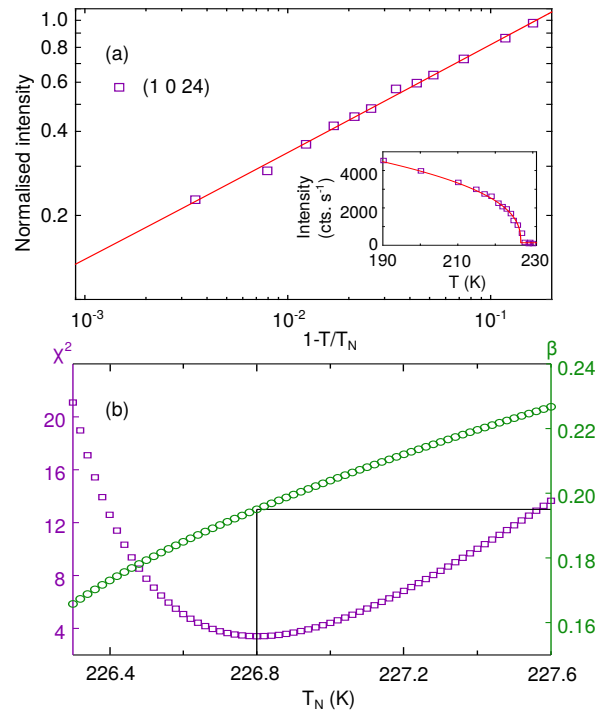}
\caption{\label{fig_orderp}
Order parameters of the magnetic critical scattering in \srsl: (a) integrated intensity and power law fit (solid line) corresponding to best-fit values of $T_{\text{N}}$=226.80(6) K and $\beta$=0.195(4). (b): Goodness of fit $\chi^2$ (purple squares) and resulting exponent $\beta$ (green circles) as function of chosen ordering temperature $T_N$. The fit in (a) corresponds to the minimum in $\chi^2$.}
\end{figure}

Here we report a comprehensive XRMS study of the critical magnetic fluctuations in \srsl\ for temperatures both below and above $T_{\text{N}}$. Our results provide decisive evidence for the importance of $XY$ anisotropy in the exchange interactions for \srsl. Below $T_{\text{N}}$, accurate values are obtained for the order parameter $\beta$ revealing that \srsl\ fits into the 2D $XY$ universality class with $h_4$ anisotropy. Above $T_{\text{N}}$ we greatly extend the temperature window over which the critical scattering has been measured to $T_{\text{N}}$ + 73~K, and carefully apply corrections for the finite instrumental resolution. We establish that in contrast to an earlier experimental study for \srsl\ \cite{fujiyama2012} the intrinsic isospin fluctuations for $T>T_{\text{N}}$ + 5~K  are dominated by $XY$ type anisotropy.

The critical scattering experiments were performed on beamline I16, Diamond Light Source, UK, and P09, PETRA III, DESY, Germany. These experiments exploited the large enhancement of the X-ray resonant magnetic scattering cross-section at the Ir L$_3$ edge. The single crystals of \srsl\ used in this study (dimensions $2\!\times\!1\!\times\!0.05\,\text{mm}^3$) were flux grown from phase-pure polycrystalline \srsl\, using techniques described elsewhere \cite{li2013}, and attached to the copper sample mount of a closed-cycle refrigerator. This was in turn mounted on a six-circle diffractometer configured to operate in a vertical scattering geometry. The energy of the incident photon beam was set to 11.218 keV, just below the $L_3$ edge of iridium, a value found to maximise the intensity of the X-ray resonant magnetic scattering. The incident beam size was determined to be
200$\times$20 $\mu m^2$ (H$\times$V). The polarization of the scattered X-rays was determined by using a Au (333) crystal analyser mounted on the detector arm. The temperature was measured to a precision of $\pm$0.01~K via a thermocouple secured to the sample mount by Teflon tape. The wavevector resolution of the instrument, including the effects of sample mosaic, was determined by mapping Bragg peaks in reciprocal space and was found to be typically 1.1$\times 10^{-3}$ and 3.7$\times 10^{-3}$\AA$^{-1}$ perpendicular and parallel to $\mathbf{Q}$ in the scattering plane respectively, and 1.4$\times 10^{-3}$\AA$^{-1}$ out of the plane.

The first objective was to determine $T_{\text{N}}$ and $\beta$ from the temperature dependence of the magnetic peak intensity $I_M$ below $T_{\text{N}}$. The results are summarized in Fig.\ \ref{fig_orderp}.  A fit to the simple power-law form $I_M\propto [(T_{\text{N}}-T)/T_{\text{N}}]^{2\beta}$ is shown in Fig.\ \ref{fig_orderp}(a) which yields $\beta$ = 0.195(4). This value is consistent with that provided by neutron scattering measurements \cite{dhital2013, ye2013}. The value of $\beta$ deduced by this analysis deviates significantly from theoretical values for both 2D Ising $(\beta=1/8)$ and 3D systems $(\beta \sim 0.35)$, but rather is consistent with value for the 2D $XYh_4$ universality class in the strong anisotropy limit \cite{taroni2008}.

Next, a detailed investigation of the critical scattering above $T_{\text{N}}$ was undertaken. For these studies the high photon energy and relatively broad energy resolution of X-ray diffractometers ($\gtrsim$ 1 eV) offer an advantage over neutrons in that they provide an accurate frequency integration to yield the instantaneous magnetic scattering function S($\mathbf{Q}$).  On the other hand the intrinsic high wavevector resolution of X-ray techniques presents a challenge in terms of following weak critical magnetic scattering to high temperatures as it broadens and weakens further. X-ray experiments to determine the magnetic critical scattering above $T_{\text{N}}$ have revealed that the critical fluctuations just above $T_{\text{N}}$ have two components: a sharp ``central'' peak, typically with a Lorentzian squared lineshape, believed to be an extrinsic feature due to the presence of defects; and a broader, weaker peak with a Lorentzian line shape arising from intrinsic critical fluctuations \cite{collins1989, thurston1994}. Realisation that the magnetic critical scattering above $T_{\text{N}}$ can have two such components mirrors earlier results for structural phase transitions \cite{andrews1986, mcmorrow1990, huennefeld2002}.

\begin{figure}
\includegraphics[width=1\columnwidth]{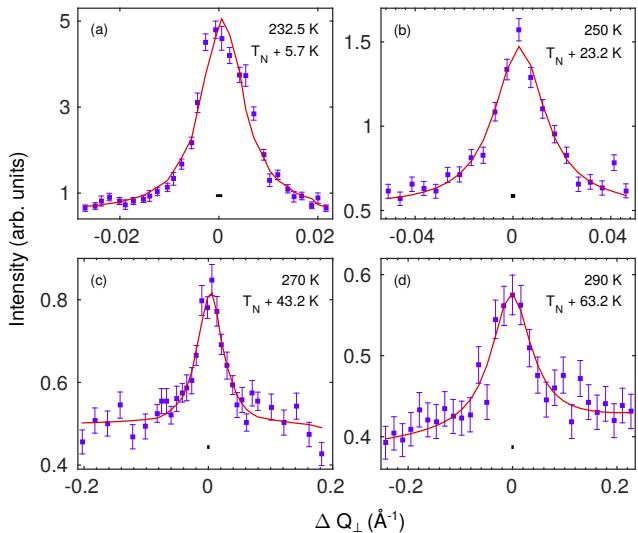}
\caption{\label{fig_scans} Critical scattering data for $T>T_{\text{N}}$
obtained in scans performed perpendicular to the wavevector transfer, $\mathbf Q$, of the
appropriate antiferromagnetic Bragg peak.
Solid lines: fits of a Lorentzian line shape convoluted with the experimental resolution. The horizontal bars under the peaks represent the resolution full width at half maximum.
}
\end{figure}

Representative scans of the magnetic critical scattering above $T_{\text{N}}$ are shown in Fig.\ \ref{fig_scans}. 
The critical scattering could be followed out to $T_{\text{N}}$ + 73~K, roughly trebling the region probed above $T_{\text{N}}$ in Ref. \cite{fujiyama2012}, 
which, as we shall show, places a much tighter constraint on the relevant form of the effective Hamiltonian.  
The critical scattering above $T_{\text{N}}$ was fitted with several different lineshapes convoluted with 
the instrumental resolution function. The most satisfactory over most of the temperature range was a Lorentzian, 
as expected for intrinsic magnetic critical scattering. 
However for  $T_{\text{N}}  < T  < T_{\text{N}}$ + 5.5~K a sharper, second component appeared to develop, presumably due to extrinsic, defect mediated scattering.
As it proved difficult to obtain unique fits in this temperature interval we do not further consider the data 
taken close to $T_{\text{N}}$.

\begin{figure}[t!]
\includegraphics[width=1\columnwidth]{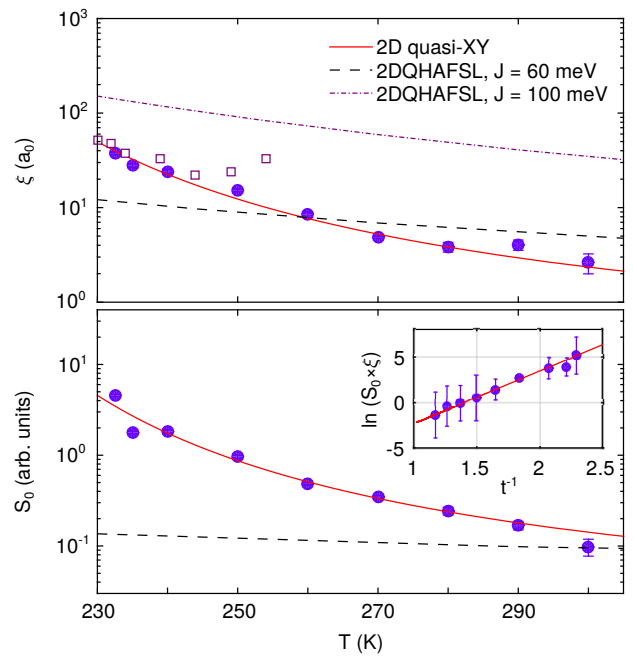}
\caption{\label{fig_214} Temperature dependence of the correlation length (top panel) and the amplitude (bottom panel) of the critical scattering for \srsl\ as filled circles. Open squares are data taken from Fujiyama \cite{fujiyama2012} divided by $a_0$. QMC results with $J$ = 60 meV (100 meV) as a dashed line (dot-dashed line) Ref.~\cite{makivic1991}.
Solid line: fit of 2D model with $XY$ anisotropy as discussed in text. Inset: Linear scaling of $\ln (\xi S_0) \sim \left(3-\eta\right)bt^{-1}$.}
\end{figure}

In Fig.\ \ref{fig_214} we compare our data with with various theories and earlier experimental results \footnote{Furthermore we believe that the values previously reported for the in-plane correlation length are too large by a factor of $a_0$. We base this on comparisons of our raw experimental data with those presented in Ref.~\cite{fujiyama2012}; the half widths at half maximum are comparable, showing that the two data sets are consistent.}. The broken lines
represent QMC results for the 2DQHAFSL model for which $\xi(T)=0.276\,a_0 e^{1.25J/k_BT}$ and $S_0\propto\xi^2T^2$, where $a_0=3.9$~\AA{} is the Ir-Ir nearest neighbour distance.
It is evident that in the 2D regime the values of $\xi$ and $S_0$ of the intrinsic Lorentzian component have much stronger temperature dependences than the expectations of the 2DQHAFSL. The 2DQHAFSL model cannot therefore describe the critical fluctuations in 
\srsl\ whatever value of $J$ is chosen. This important realisation disagrees with the main conclusion of Ref.~\cite{fujiyama2012}.

We therefore considered alternative models. 
In the 2D $XY$ model a phase transition occurs at the Kosterlitz-Thouless temperature, $T_{\text{KT}}$ \cite{kosterlitz1973}. Above $T_{\text{KT}}$ the correlation length $\xi$ and scattering amplitude $S_0$ are described by: $\xi \propto \exp{\left(b/\sqrt{t}\right)}$ and $S_0 \propto \,\exp{\left[b\,(2-\eta)/\sqrt{t}\right]},$
where $b$ is a non-universal constant (typically $ b\!\approx\!1.9$), $\eta\!=\!1/4$ and $t\!=\!\left(T -T_{\text{KT}}\right)/\,T_{\text{KT}}$ the reduced temperature  \cite{bramwell1993}.
For real systems, three-dimensional order at a temperature $T_{\text{N}} > T_{\text{KT}}$ occurs as a result of weak interlayer coupling 
$J^\prime$ amplified by the diverging size of correlated regions $\xi(T)^2$. However attempts to fit the pure 2D $XY$ model to our data yielded unphysical values of $b=3.0(7)$ and $J'/J=5\!\times\!10^{-6}$, and were thus discarded.

Instead, a consistent description of our data was obtained using the 2D anisotropic Heisenberg model (2DAH), where leading Heisenberg interactions are augmented by easy-plane, $XY$ anisotropy \cite{cuccoli2003_prl, *cuccoli2003_prb}. The 2DAH model also has a 
correlation length that diverges exponentially towards a finite temperature $T_{\text{KT}}$, placing it in the 2D $XY$ universality class.
In the limit of long correlation lengths, $\xi$ and $S_0$ scale the same way as for the pure 2D $XY$ model. However with increasing temperature there is a crossover towards $\xi(T)=\xi_0\exp{(b/t)}$ and $S_0\propto \exp{\left[b\,(2-\eta)/t\right]}$ for $\xi \lesssim 100\,a_0$ \cite{cuccoli2003_prl, *cuccoli2003_prb}. Our data fall in this regime.
 
The solid lines in Fig.\ \ref{fig_214} show the fit to the 2DAH which is seen to provide an excellent description of the data in the 2D regime. From the fits we obtain $\xi_0\!=\!0.9(1)$~\AA, $T_{\text{KT}}\!=\!162(10)$~K and $b\!=\!2.1(7)$, with the latter in good agreement with theory and experiment for other layered materials \cite{bramwell1993, alsnielsen1993, ronnow2000}. 
From the value of the 2D correlation length at $T_N$ it is possible to estimate the ratio of the in-plane and inter-layer couplings $J'/J \simeq \xi(T_N)^{-2} = \exp(-2b/t_N) \approx0.001$.
This ratio is an order of magnitude higher than for the cuprates \cite{manousakis1991, keimer1992}, which can be understood by comparing the shapes of the ground state orbitals of \srsl\ and \lacuo. The $j_{\text{eff}}$=1/2 state is approximately cubic, whereas for \lacuo\ the d$_{\text{x}^\text{2}-\text{y}^\text{2}}$ orbital lies predominantly in the \emph{xy} plane. Thus larger out-of-plane interactions would be expected for \srsl\ compared with \lacuo. The description of $\xi(T)$ by inclusion of $XY$ anisotropy is consistent with the extracted 
value of $\beta=0.195(4)$, which places the fluctuations below $T_N$ in the 2D $XYh_4$ universality window, where a 4-fold in-plane anisotropy pushes $\beta$ from 0.231 for 2D $XY$ towards $\beta=1/8$ for 2D Ising \cite{taroni2008} interactions.

In general, the effect of $XY$ anisotropy on the spin wave dispersion is to lift the degeneracy between the Heisenberg modes. The in-plane fluctuations $\omega_{\parallel}$ maintain a continuous symmetry and are hence gapless (Goldstone mode), whereas the out-of-plane fluctuations $\omega_{\perp}$ are gapped. At the zone centre the out-of-plane spin wave gap is given by 
${\omega_0 = 4JSZ_c\sqrt{2\Delta_{\lambda}}}$, where $0\!\leq\!\Delta_{\lambda}\!\leq\!1$ parametrizes the easy-plane anisotropy \cite{greven1995}. In data from a recent high-resolution RIXS experiment on Sr$_2$IrO$_4$ \cite{kim2014}, there appears to be a magnon gap of $\sim$30 meV at the magnetic zone centre, qualitatively consistent with the existence of  $XY$ anisotropy.
The degree of $XY$ anisotropy required to produce a gap of this size may be determined by extending the model of Kim {\it et al.} \cite{kim2014} to include 
anisotropic exchange. The resulting 2DAH Hamiltonian reads
\begin{align}
\label{Ham}
\mathcal{H}_{ab}&=\sum_{\langle i,j \rangle} \tilde{J} \left[S_i^xS_j^x + S_i^yS_j^y + \left(1-\Delta_{\lambda}\right)S_i^zS_j^z \right] \nonumber\\
&+ \sum_{\langle\langle i,j \rangle\rangle} J_2\vec{S}_i\vec{S}_j + \sum_{\langle\langle\langle i,j \rangle\rangle\rangle} \!\!\!J_3\vec{S}_i\vec{S}_j,
\end{align}
Here $\tilde{J} = J_{iso}/(1-\Delta_{\lambda})$ is the effective nearest-neighbour (nn) exchange parameter, $J_{iso}$ is the isotropic Heisenberg nn exchange, and $J_2,J_3$ symbolizes the exchange between next-nearest and third nearest neighbours respectively. Out-of-plane exchange coupling has been neglected since as demonstrated above, it is a factor of $\sim\!10^3$ weaker than the in-plane terms, and thus not resolvable with RIXS at present. 

\begin{figure}[h!!]
\includegraphics{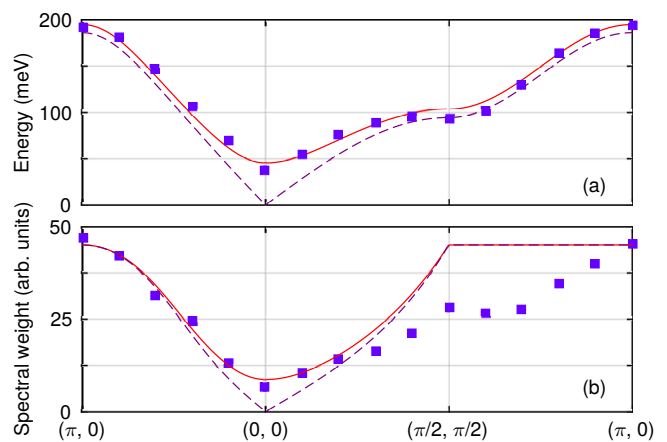}
\caption{Fit of the anisotropic 2D Heisenberg model to the experimental single magnon dispersion. Squares: Energy (a) and spectral weight (b) of the single magnon peak obtained from fitting RIXS data provided in Ref.~\cite{kim2014}. Errorbars are smaller than the displayed symbols. Solid and dashed lines: best fit of gapped and gapless modes to experimental data with $\tilde{J}=57(7)$ meV, $J_2=-18(3)$ meV, $J_3=14(2)$ meV, $\Delta_{\lambda} = 0.08(1)$.}
\label{disp}
\end{figure}

We fitted the theoretical dispersion of this 2DAH model to the experimental data from Ref.~\cite{kim2014} using a linear spin-wave model derived from Eq.~\ref{Ham} with the results shown in 
Fig.\ \ref{disp}. There is good agreement between the experimental dispersion and that calculated from the anisotropic model, with the fitted exchange constants essentially identical to those obtained from the 
earlier low-resolution study within the error bars \cite{kim2012_sr214}. The agreement for the spectral weight is perhaps less compelling but can be said to be qualitatively reasonable, in particular the
model reproduces the non-zero spectral weight at the zone centre. Moreover, it proved difficult to extract the spectral weight from the RIXS data \cite{kim2014} as there appears to be a significant variation of 
lineshape as a function of momentum transfer, which may be evidence for an additional excitation mode between the single magnon and bi-magnon peaks (the latter was accounted for in the fitting procedure). Fitting the dispersion and intensities to the localized spin model proposed by Igarashi \cite{igarashi2014} provided similar results, corroborating their prediction of an out-of-plane gap for \srsl. The easy-plane anisotropy parameter $\Delta_{\lambda} = 0.08(1)$ determined here should be considered as an upper bound to the true anisotropy, since the finite momentum and energy resolution means that the excitation energy at the zone centre cannot be fully resolved. Nevertheless this is significantly larger than that observed for \lacuo\ $(\Delta_{\lambda}=2.0(5)\times 10^{-4})$ \cite{keimer1993}, which further illustrates the relative importance of $XY$ anisotropy for \srsl.

In conclusion, we have presented a detailed study of the critical magnetic properties of the single-layer perovskite iridate \srsl.
Our results establish the importance of $XY$ anisotropy in its effective low-energy Hamiltonian.
Although in their seminal paper Jackeli and Khaliullin \cite{jackeli2009} proposed that the interactions in \srsl\ are isotropic (Heisenberg)  
they note that the leading anisotropy term should have $XY$ character.  Here we have shown that the $XY$ anisotropy is sufficiently strong to push the critical properties of \srsl\ significantly away from the isotropic 2DQHAFSL model.
Thus while \srsl\ and \lacuo\ share some similarities, as far as their effective magnetic interactions and critical properties are concerned, they are quite distinct.

\begin{acknowledgments}
Research in London and Oxford was supported by the EPSRC. Work in Switzerland was supported by the Swiss National Science Foundation and its Sinergia network Mott Physics Beyond the Heisenberg Model. Parts of this research were carried out at the light source PETRA III at DESY, a member of the Helmholtz-Gemeinschaft. J.~Vale would like to thank UCL and EPFL for support through an UCL Impact Studentship. We also thank S.T.~Bramwell, A.~Chubukov and G.~Khaliullin for useful discussions.
\end{acknowledgments}

\bibliographystyle{apsrev4-1}
\bibliography{references}

\end{document}